\documentstyle[aps,epsf,prl]{revtex}
\begin{document}
%\draft
\title{First-Order Transition and Critical End-Point in Vortex
Liquids in Layered Superconductors
}
\author{Joonhyun {\sc Yeo}\footnote
{E-mail : jhyeo@konkuk.ac.kr}} 
\address{Department of Physics, Konkuk University,
Seoul, Korea 143-701}
\author{M.\  A.\ {\sc Moore}}
\address{Department of Physics, University of Manchester,
Manchester M13 9PL, United Kingdom} 
\date{\today}
\maketitle

\begin{abstract}
We calculate various thermodynamic quantities
of vortex liquids in a layered superconductor by using 
the nonperturbative parquet approximation method, which was 
previously
used to study the effect of thermal fluctuations in 
two-dimensional
vortex systems. We find there is a first-order transition
between two vortex liquid phases which differ in the magnitude of 
their
correlation lengths.
As the coupling between the layers increases,
the first-order transition line ends
at a critical point. We discuss the possible relation between
this critical end-point and
the disappearance of the first-order transition which is observed 
in 
experiments on high temperature superconductors at low magnetic 
fields. 
\end{abstract}
\pacs{PACS numbers: 74.20.De, 74.60.-w}
%%%%%%%%%%%%%%%%%%%%%%%%%%%%%%%%%%%%%

\vspace{1cm}
%\twocolumn
%\narrowtext

\section{Introduction}

Thermal fluctuations and quenched disorder are two important 
factors
responsible for the rich phase diagram exhibited by high 
temperature
superconductors in a magnetic field \cite{blatter}. 
Recent experiments \cite{exp,exp1} on high temperature 
superconductors
show a first-order transition line well below the upper critical 
field
$H_{c2} (T)$ in the field-temperature (H-T) phase diagram. 
When the strength of disorder is weak, this line
is usually interpreted as a melting line of a vortex lattice into 
a vortex 
liquid.  The first-order
transition disappears at both high and low magnetic 
fields \cite{exp2,exp3,exp4,exp5}.
In the vortex lattice melting picture the loss of first-order 
behavior
at high fields is
usually attributed to the effect of disorder, which is supposed 
somehow to
produce a multi-critical point
where the first-order transition changes into a continuous one.

In a recent numerical simulation \cite{km}
of a vortex system in a layered superconductor, a 
first-order transition line was also obtained
which disappeared  at a critical end-point
at {\it low}  magnetic fields.  According to the simulation 
results,
in contrast to the vortex lattice melting
picture, there is only one phase below and above the
transition, namely the vortex liquid but with different 
correlation lengths on
either side of the first-order transition line.
However, in other simulations \cite{period} using periodic 
boundary conditions
within a layer, an apparent first-order transition between vortex 
liquid and
lattice was obtained.

In this paper we apply an analytical approach to a layered 
superconductor 
in a magnetic field perpendicular to the layers in an attempt to 
elucidate the nature of phase transitions in the system.
We use the parquet approximation \cite{ym}, which has been 
successfully applied
to the two-dimensional vortex system. It is a nonperturbative 
analytic 
method free from any finite size or boundary effects
in the direction perpendicular to
the magnetic field. The parquet approximation
deals with the renormalized four-point function of
the vortex system which is obtained by 
summing an infinite subset of Feynman diagrams,
the so-called parquet diagrams. In two dimensions this method 
is able to capture 
the growing crystalline order developing in the vortex
liquid as the temperature is lowered \cite{ym}. 
As explained in the discussion below, the introduction
of an additional dimension to the parquet calculations 
imposes severe numerical difficulties because of the 
large number of variables specifying the renormalized 
vertex function. 
This problem has restricted us in this paper 
to  considering  a small system which consists of just 
four layers satisfying  a periodic boundary condition along
 the field direction.
Each layer is, however, treated as an infinite plane so we can are 
in effect
descibing an infinite number of vortices.
Despite the unphysically small size of the system
we consider here, we find that the effect of the inter-layer
couplng makes a significant difference compared with the 
two-dimensional
case, where no finite-temperature phase transition occurs 
in the parquet calculation. In the layered system, we  
find that the thermodynamic quantities 
describing the vortex liquid show abrupt changes
when the field and the temperature are varied, and these sharp  
changes
 are interpreted as
a first-order phase transition within the vortex liquid phase. We 
find that
the first-order transition line ends at a critical end-point.
The two phases we obtain above and below
the transition are both liquid phases with different 
correlation lengths, 
consistent with the numerical simulation results \cite{km}.

In the next section we present our model for the vortex liquid
in a layered superconductor. We then set up the parquet equations 
for the layered system. In the final section we present 
the numerical solutions of the parquet equation for the four-layer
system which are interpreted in terms of a first-order 
transition and the termination of the first-order
transition line at a critical end-point. We conclude with a 
discussion on the 
possible implication of our 
results for the 
H-T phase diagram of a layered superconductor.

\section{model}

Our starting point is the Lawrence-Doniach model for a layered
superconductor in a magnetic field perpendicular to the layers.
With the order parameter in the n-th layer denoted by $\psi_n$,
the free energy is given by
\begin{eqnarray}
&&F[\psi,\psi^*] = \sum_{n} d_0 \int d^2 {\bf r} 
\Big[ \alpha |\psi_n({\bf r})|^2
+\frac{\beta}{2}|\psi_n ({\bf r})|^4  \nonumber \\
&&~~~~~+\frac{1}{2m_{ab}}|(-i\hbar\nabla-\frac{e^*}{c}{\bf 
A})\psi_n|^2
\nonumber \\
&&~~~~~+\frac{\hbar^2}{2m_c d^2} |\psi_n ({\bf r}) - \psi_{n+1} 
({\bf r})|^2
\Big] , 
\end{eqnarray}
where $d_0$ is the layer thickness, $d$ the layer spacing and 
$\alpha$, $\beta$, $m_{ab}$, and $m_c$ phenomenological 
parameters.
We denote by $\tau\equiv \hbar^2/2m_c d^2 = (\xi_c /d)^2$
the dimensionless ratio between the coherence length $\xi_c$
perpendicular to the layers and the layer spacing. We take ${\bf 
B}
=\nabla\times{\bf A}$ as constant and uniform.  

We use the lowest Landau level (LLL) approximation which is 
believed 
to be valid over a large portion of the vortex liquid regime.
We expand the order parameter $\psi_n({\bf r})$ 
in terms of the eigenstates of 
$|(-i\hbar\nabla-\frac{e^*}{c}{\bf A})\psi_n|^2$ and keep
only the lowest eigenvalue state. 
In the symmetric gauge, where ${\bf A}=\frac{B}{2}(-y,x,0)$,
the LLL wavefunction is given by
$\psi^{\rm LLL}_n ({\bf r})=\exp(-\mu^2 |z|^2/4)\phi_n(z)$ where
$\mu^2 = e^* B /\hbar c$ and 
$\phi_n(z)$ is an arbitrary analytic function of $z=x+iy$.
In the LLL approximation, the free energy becomes
\begin{eqnarray}
&&F[\phi,\phi^*] = \sum_n d_0\int dz^* dz \Big[ 
\alpha_H e^{-\mu^2 |z|^2/2}
|\phi_n(z)|^2  \nonumber  \\
&&~~~~~+ \frac{\beta}{2} e^{-\mu^2 |z|^2} |\phi_n(z)|^4 \nonumber 
\\ 
&&~~~~~+\tau e^{-\mu^2 |z|^2/2} 
|\phi_{n}(z)-\phi_{n+1}(z)|^2\Big], 
\end{eqnarray}
where $\alpha_H \equiv \alpha + e^* B \hbar /2c m_{ab} $ changes 
sign
crossing the upper critical field line $H_{c2} (T)$. 
Physical properties
of the vortex system in the layered superconductor are then 
determined
from the partition function
$
Z=\int \prod_n {\cal D}\phi_n {\cal D}\phi^*_n \exp 
(-F[\phi,\phi^*]
/k_B T) . 
$

For a two-dimensional system,
that is for a single layer, the temperature and field dependence  
are all 
contained within the single dimensionless parameter $\alpha_{2T}
\equiv(2\pi d_0 /\beta\mu^2 k_B T)^{1/2} \alpha_H$. 
The inter-layer coupling 
strength is described by the dimensionless parameter 
$\tau_T\equiv(2\pi d_0 /\beta\mu^2 k_B T)^{1/2}
\tau$. For the layered system it is very convenient
to use the dimensionless field-temperature parameter
$\alpha_T$ employed in studies of
the continuous anisotropic three dimensional GL model:
$\alpha_T=(8\pi\hbar c /\beta^\prime e^* B k_B T)^{2/3} 
(\hbar^2/2m_c)^{1/3} \alpha_H$ where $\beta^\prime =(d/d_0)\beta$ 
is the coefficient of the quartic term for the three dimensional
GL model. The two-dimensional limit corresponds to 
$\tau_T\rightarrow 0$,
while in the limit $\tau_T\rightarrow\infty$ for constant 
$\alpha_T$,
the system behaves as a continuous three dimensional model.
Note that $\alpha_T = 2^{4/3}\tau_T^{1/3} \alpha_{2T}$. 
In terms of $t\equiv T/T_{c0}$ and $h\equiv H/H_{c2}(0)$,
\begin{equation}
\alpha_T \sim \frac{1-t-h}{(th)^{2/3}}, ~~~~~\tau_T \sim 
\frac{1}{(th)^{1/2}},
\label{attt}
\end{equation}
where $T_{c0}$ is the zero-field transition temperature and 
$H_{c2}(0)$ is
 the straight line extrapolation 
of the $H_{c2}(T)$ line near $T_{c0}$ to zero temperature.

We shall calculate various correlation functions using the
parquet approximation. This is a nonperturbative
analytic approximation and requires no boundary conditions to be 
imposed
in the direction perpendicular to the magnetic field within each 
layer. 
Our system consists of a stack of $N$ layers,
on which we impose the periodic boundary condition in the 
direction
along the field
$\phi_{n+N}(z)=\phi_n (z)$. We introduce the Fourier transform
$\widetilde{\phi}_q$ of $\phi_n$ via
\begin{equation}
\phi_n(z)=\frac{1}{dN}\sum_{q\in 1st. B. Z.} 
e^{iqnd}\widetilde{\phi}_q (z), 
%\widetilde{\phi}_q (z) = d\sum_{n=1}^N \phi_n (z) e^{-iqnd},
\end{equation} 
where $q=2\pi j /Nd$ and we use $N$ integer values of $j$ in
the sum such that the wavevector $q\in [-\pi/d ,\pi/d)$ belongs to 
the first
Brillouin zone.

One can develop the standard perturbation theory 
from the given partition function.
The bare propagators are given by
\[
\langle\widetilde{\phi}^*_{q^\prime} (z^{\prime *}) 
\widetilde{\phi}_q
(z)\rangle_0 = d^2 N\delta_{q,q^\prime}(\mu^2/2\pi)
e^{\mu^2 z^{\prime *}z/2} G_0 (q), 
\]
where 
\begin{equation}
G_0 (q)=\Big(\frac{k_B T}{d_0}\Big)
\frac{1}{\alpha_H + 2\tau (1-\cos(qd))}.
\label{prop}
\end{equation}
Since the magnetic length $\mu^{-1}$ is the only length scale
perpendicular to the field direction which appears 
in the propagator \cite{ym},
the fully renormalized propagator can also be written as 
(\ref{prop})
with the renormalized $G(q)$ replacing the bare function $G_0(q)$.
The renormalized $G(q)$ is determined self-consistently in the
parquet approximation.

The main quantity one studies in the parquet approximation
is the renormalized connected four-point function,
$
\langle\widetilde{\phi}^*_{q_1}(z_1^*)\widetilde{\phi}^*_{q_2}
(z_2^*)\widetilde{\phi}_{q_3}(z_3)\widetilde{\phi}_{q_4}(z_4)
\rangle_c
$ 
which can be written in terms of
the renormalized 
vertex function $
\Gamma (q_1,q_2,q_3;{\bf k})=\Gamma (q_1, q_2, q_3, q_1 + q_2 - 
q_3;
{\bf k})$ \cite{ym}.
Here the wavevector  ${\bf k}$ lies in the two dimensional space
perpendicular to the magnetic field and $q_i$ is a wavevector 
along the
field direction.  
Note that to the lowest
order $\Gamma (q_1,q_2,q_3;{\bf k})=
\Gamma_B({\bf k})$ independent of the wavevectors along the field 
direction. 

In order to make a resummation over all parquet diagrams, we note 
that
the contributions to $\Gamma$ can be decomposed into a totally 
irreducible part
and a reducible part which in turn can be written
as the sum of three parts $\Pi_i (i=1,2,3)$ 
representing the contributions 
from three different channels. 
(A detailed discussion on the diagrammatic decomposition can
be found in Ref.\cite{ym}).  
The parquet approximation we employ here corresponds
to neglecting in the totally irreducible
vertex all the higher order (O($\beta^4$)) diagrams except
the bare vertex function $\Gamma_B ({\bf k})$ so
\begin{equation}
\Gamma(q_1,q_2,q_3;{\bf k})=\Gamma_B ({\bf k})
+\sum_{i=1}^3 \Pi_i (q_1,q_2,q_3;{\bf k}).
\label{o1}
\end{equation} 
Each reducible vertex $\Pi_i$ is composed of the irreducible 
vertex
$\Lambda_i$ where  
\begin{equation}
\Lambda_i (q_1,q_2,q_3;{\bf k})=\Gamma_B({\bf k}) +
\sum_{j\neq i}\Pi_j(q_1,q_2,q_3;{\bf k}) \label{o2}
\end{equation} 
and the renormalized $\Gamma$ via the following Bethe-Salpeter 
equations:
%\onecolumn
%\widetext
\begin{eqnarray}
&&\Pi_1 (q_1,q_2,q_3;{\bf k})=
-\frac{1}{N}\sum_p \widehat{G}(p)\widehat{G}(q_1 +q_2 -p) 
\Big[\Lambda_1 (q_1,q_2,p)\circ
\Gamma(p,q_1 + q_2 -p,q_3)\Big] ({\bf k}), \nonumber \\
&&\Pi_2 (q_1,q_2,q_3;{\bf k})=
-\frac{2}{N}\sum_p \widehat{G}(p)\widehat{G}(p-q_1+q_3) 
\Lambda_2 (q_1,p-q_1+q_3,q_3;{\bf k})
\Gamma (p,q_2,p+q_3-q_1;{\bf k}), \nonumber \\
&&\Pi_3 (q_1,q_2,q_3;{\bf k})=
-\frac{2}{N}\sum_p \widehat{G}(p)\widehat{G}(p+q_2-q_3) 
\Big[\Lambda_3 (q_1,p+q_2-q_3,p)
\ast\Gamma(p,q_2,q_3)\Big]({\bf k}), \label{o3} 
\end{eqnarray}
where the operation $\circ$ 
between two arbitrary functions
$f({\bf k})$ and $g({\bf k})$ is defined by
%\narrowtext
\begin{eqnarray*}
(f\circ g)({\bf k})&=&\frac{2\pi}{\mu^2}\int\frac{
d^2 {\bf k}^\prime}{(2\pi)^2}\; f({\bf k}-{\bf k}^\prime)
g({\bf k}^\prime) \\
&&\times\cos((k_x k^\prime_y -k_y k^\prime_x)/\mu^2),
\end{eqnarray*}
and $f\ast g$ is just the convolution which is the same as 
the above expression without the cosine term.
In (\ref{o3}) we have used the dimensionless form of the 
propagator function $\widehat{G}(q)\equiv (d_0 \beta \mu^2 /2\pi
k_B T)^{1/2} G(q)$. 
The above equations, (\ref{o1}), (\ref{o2}) and (\ref{o3}) form
a closed set for $\Gamma$ when 
the renormalized propagator
$\widehat{G}(q)$ is known. 
In the parquet approximation $\widehat{G}(q)$ is determined 
self-consistently by the Dyson equation:
%\widetext
\begin{eqnarray}
&&\widehat{G}^{-1}(q)-
\widehat{G}^{-1}_0 (q)=
\frac{2}{N}\sum_{q^\prime} \widehat{G}(q^\prime) \nonumber \\
&&~~~~~~~~~~~~~~~-\frac{2}{N^2}
\sum_{q^\prime,q^{\prime\prime}}\widehat{G}(q^\prime)
\widehat{G}(q^{\prime\prime})
\widehat{G}(q+q^\prime-q^{\prime\prime})\frac{2\pi}{\mu^2}\int
\frac{d^2 {\bf k}}{(2\pi)^2}\Gamma(q,q^\prime,q^{\prime\prime};
{\bf k})e^{-{\bf k}^2/2\mu^2}. \label{do}
\end{eqnarray} 

%\narrowtext
Using the solutions to the above equations one can calculate 
several
interesting physical quantities. The structure factor, which
is the measure of correlation between vortices in the vortex
liquid, is calculated from 
\begin{eqnarray*}
\chi_{n-n^\prime}({\bf r}-{\bf r}^\prime)& =& \langle
|\Psi_n ({\bf r})|^2 |\Psi_{n^\prime} ({\bf r}^\prime)|^2\rangle
\nonumber \\
&&-\langle|\Psi_n ({\bf r})|^2 \rangle\langle
\Psi_{n^\prime} ({\bf r}^\prime)|^2\rangle .
\end{eqnarray*}
The structure factor $\Delta_m ({\bf k})$
used in this paper is then defined by 
\begin{equation}
\Delta_m ({\bf k}) \equiv 
\Big(\frac{d_0 \beta}{k_B T}\Big) e^{{\bf k}^2/2\mu^2}
\int d^2 {\bf R} e^{i {\bf k}\cdot{\bf R}} \chi_m ({\bf R}). 
\label{structure}
\end{equation}
Using (\ref{do}), one can write the generalized
Abrikosov ratio $\beta_A$ as 
\begin{equation}
\beta_A \equiv \frac{\langle|\Psi_n ({\bf r})|^4\rangle}
{\langle|\Psi_n ({\bf r})|^2\rangle^2} = 
\frac{1 - N^{-1}\sum_q \widehat{G} (q) \widehat{G}^{-1}_0 (q)}
{\Big[N^{-1}\sum_q \widehat{G}(q)\Big]^2}.
\label{betaa}
\end{equation}
The Josephson coupling parameter $\eta$ measures the degree of 
independence between adjacent layers and is given by
\begin{equation}
\eta\equiv\frac{\langle|\Psi_n({\bf r})-\Psi_{n+1}
({\bf r})|^2\rangle}{\langle|\Psi_n ({\bf r})|^2\rangle}
=\frac{N^{-1}\sum_q 2(1-\cos (qd))\widehat{G}(q)}
{N^{-1}\sum_q \widehat{G}(q)}
\label{eta}
\end{equation}

We can also put the above equations in more convenient form as 
follows. 
If we introduce
%\widetext 
\begin{equation}
\widetilde{\gamma}(q_1 , q_2, q_3;{\bf k})\equiv \Big[ 
\widehat{G}(q_1)
\widehat{G}(q_2) \widehat{G}(q_3) \widehat{G}(q_1+q_2-q_3)
\Big]^{1/2}\Gamma(q_1, q_2, q_3;{\bf k}). 
\end{equation}
and similary $\widetilde{\lambda}_i(q_1,q_2,q_3;{\bf k})$
and $\widetilde{\pi}_i(q_1, q_2, q_3;{\bf k})$ from 
$\Lambda_i(q_1,q_2,q_3;
{\bf k})$ and $\Pi_i(q_1,q_2,q_3;{\bf k})$ respectively, and
denote the Fourier transform of $\widetilde{\gamma}$, 
$\widetilde{\pi}$ and $\widetilde{\lambda}$ by
$\gamma(\{n_i\};{\bf k})$, $\pi_i(\{n_i\};{\bf k})$ and
$\lambda_i(\{n_i\};{\bf k})$ with the shorthand notation
$\{n_i\}=(n_1,n_2,n_3,n_4)$,
then (\ref{o3}) is simplified to
%\widetext
\begin{eqnarray}
&&\pi_1 (\{n_i\};{\bf k})=
-\sum_{n^\prime,n^{\prime\prime}}  
\Big[\lambda_1 (n_1,n_2,n^\prime ,n^{\prime\prime})\circ
\gamma(n^\prime ,n^{\prime\prime},n_3,n_4)\Big] ({\bf k}) 
\nonumber \\
&&\pi_2 (\{n_i\};{\bf k})=
-2\sum_{n^\prime,n^{\prime\prime}}
\lambda_2 (n_1,n^\prime,n_3,n^{\prime\prime};{\bf k})
\gamma(n^{\prime\prime},n_2,n^\prime,n_4;{\bf k}) \nonumber \\
&&\pi_3 (\{n_i\};{\bf k})=
-2\sum_{n^\prime,n^{\prime\prime}} 
\Big[\lambda_3 (n_1,n^\prime,n^{\prime\prime},n_4)
\ast\gamma(n^{\prime\prime},n_2,n_3,n^\prime)\Big]({\bf k}).
\label{ooo3}
\end{eqnarray}
%\narrowtext
The remaining parquet equations can easily be derived as
follows:
\begin{equation}
\lambda_j(\{n_i\} ;{\bf k})=\gamma(\{n_i\} ;{\bf k})
-\pi_j(\{n_i\} ;{\bf k}) \label{ooo2}
\end{equation}
for $j=1,2,3$ and
\begin{equation}
\gamma (\{n_i\} ; {\bf k}) = f_B (\{n_i \} )\Gamma_B ({\bf k})
+\sum_{j=1}^3 \pi_j (\{n_i\} ;{\bf k}), \label{ooo1}
\end{equation}
where
\begin{eqnarray}
f_B (\{n_i\})&=&\sum_l {\cal G}(l-n_1){\cal G}(l-n_2) \nonumber \\
&&\times{\cal G}(n_3-l){\cal G}(n_4-l)
\end{eqnarray}
with ${\cal G}(n)$ being the Fourier transform of 
$\widehat{G}^{1/2}
(q)$: ${\cal G}(n)=(1/N)\sum_q \exp (iqnd)\widehat{G}^{1/2}(q)$.

\section{Results}

We solved (numerically!) the parquet equations, 
(\ref{ooo3})-(\ref{ooo1})
for $\gamma(\{n_i \};{\bf k})$ with Eq.~(\ref{do}) for 
the propagator $G(q)$. One could use equivalently the set of 
equations,
(\ref{o1})-(\ref{o3}) for $\Gamma(\{q_i\};{\bf k})$.
It is just a matter of convenience. In the present analysis 
the former was used. 
In solving we start from some
initial functions, $\lambda_i$, $\gamma$ and $G(q)$ and
update these functions iteratively using
(\ref{ooo3})-(\ref{ooo1}) and (\ref{do}). 
As mentioned above, the main numerical difficulty compared to the 
two-dimensional case 
is that $\gamma$ and
$\lambda_i$ now involve three extra indices
in addition to the two-dimensional momentum ${\bf k}$.
A large amount of computer
memory and CPU time is required
as the number of layers becomes large.
In this paper we only consider a small system 
where the number of layers is four and try to see what effect the
coupling between the layers has on the two-dimensional parquet 
results.
At high temperatures we find that the iteration 
converges very quickly, but as the temperature is lowered the 
convergence gets slower, and furthermore we
have to use the results obtained at a nearby temperature as
the initial values for $\lambda_i$ and $\gamma$ in order to get 
rapid
convergence. 
 
From the solutions
$\Gamma$ and $G(q)$ to the parquet equations 
for given $\alpha_{2T}$ and $\tau_T$, we calculated the 
thermodynamic
quantities introduced in the previous section.
The three-dimensional temperature parameter $\alpha_T$
can be obtained from the two parameters.
Figures 1 and 2 show the Abrikosov ratio $\beta_A$, defined in
Eq.~(\ref{betaa})
and the Josephson coupling parameter $\eta$ in Eq.~(\ref{eta}) as
functions of $\alpha_T$
for various values of the interlayer coupling $\tau_T$. We
find that for some values of $\tau_T$,
these thermodynamic quantities show abrupt changes 
as $\alpha_T$ is varied. But this behavior disappears
for $\tau_T \ge 0.11$ when now 
$\beta_A$ and $\eta$ just decrease smoothly 
without showing any sudden drops as the temperature 
is lowered.
These features 
can be explained by the existence of a
first-order transition in the vortex
liquid and of a critical end-point: the first-order nature
of the transition gets weaker as the inter-layer coupling
strength $\tau_T$ increases, eventually disappearing at
the critical end-point at $\tau_T\simeq 0.11$.

\newpage 

\begin{figure}
\centerline{\epsfxsize=7cm\epsfbox{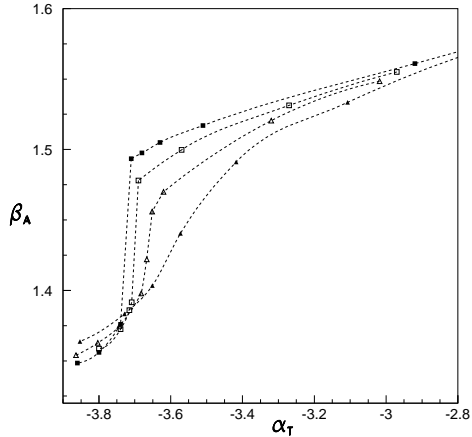}}
\vspace{10pt}
\caption{The Abrikosov ratio $\beta_A$ as a function of 
temperature
parameter $\alpha_T$  
for different inter-layer coupling strengths 
$\tau_T=0.1$ (filled squares),
0.105 (open squares), 0.11 (open triangles)
and 0.12 (filled triangles). The dotted lines are 
guides to the eye.}
\end{figure}

\begin{figure}
\centerline{\epsfxsize=7cm\epsfbox{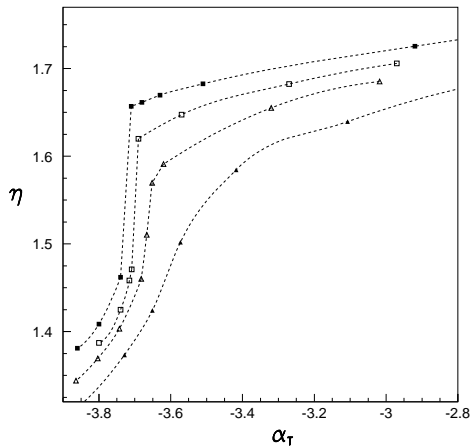}}
\vspace{10pt}
\caption{The Josephson coupling parameter $\eta$ 
as a function of temperature
parameter $\alpha_T$  
for different inter-layer coupling strengths $\tau$.
The symbols are the same as in Fig.~1.
The dotted lines are guides to the eye.}
\end{figure}

The sudden changes in these quantities occur in a very narrow 
range of the temperature parameter $\alpha_T$. We believe
it is not a crossover which just happens to resemble a first order 
transition,
since we find instabilities in obtaining the solutions
to the parquet equations within this narrow
range of $\alpha_T$. This is illustrated in Fig.~3. 
It shows a typical example of how  thermodynamic quantities
such as $\beta_A$ or $\eta$ evolve as we iteratively solve the 
parquet equations in the vicinity of the transition.
Well above the transition, $\beta_A$ 
converges rather quickly as can be seen in Fig.~3. 
This is also the case well 
below the transition. However, as the transition is approached
from above the convergence gets slower, then we reach a
very narrow region of $\alpha_T$ where the iteration appears not 
to be
converging. We find this kind of instability for the cases
where $\tau_T < 0.11$,
while no instability is seen for $\tau_T \geq 0.11$. This is 
in fact how we locate the critical end-point in our model.
Since the iteration method works well only if we use initial
functions which are close to the actual solutions,
we conclude that this instability signals a sharp change 
emerging in the system. 
Strictly speaking, there should exist a solution for any value of
$\alpha_T$ even if there is a first order transition. Therefore 
the instability does not imply that there is no convergent
solution. It just tells us that it is very hard to get a solution
using the iteration method employed  here. (It is interesting
to note that the solution below the transition was in fact 
found after a very long iteration). For this reason,
although we find that the size of the first-order transition-like 
jumps 
in thermodynamic quantities such as $\beta_A$ decreases
as the critical end-point is approached,
it is quite hard to determine accurately the size
of these jumps. 
  
\begin{figure}
\centerline{\epsfxsize=7cm\epsfbox{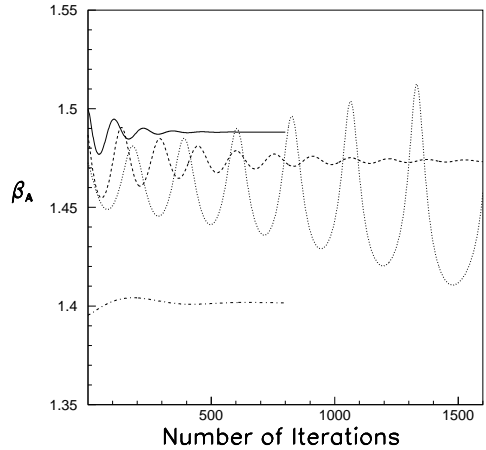}}
\vspace{10pt}
\caption{The evolution of the Abrikosov ratio
$\beta_A$ in the iteration
of the parquet equations. The inter-layer coupling is fixed at
$\tau_T=1.075$. The three dimensional 
temperature parameter $\alpha_T$ is given by
-3.594 (solid), -3.654 (dashed), -3.684 (dotted),
and -3.690 (dot dashed).
}
\end{figure}

The first order transitions we obtain here are between two vortex 
liquid 
phases with the phase below the transition being more correlated.
This can be seen in Fig.~4 where the structure factor 
$\Delta_n (K)$ defined in 
Eq.~(\ref{structure})
is shown at temperatures above
and below the transition. In both cases, there is no crystalline
long-range order present in the system. 
We note that the first peak at $K\simeq 2.69$, which is the 
value of the first reciprocal lattice vector of a triangular 
lattice, 
represents the crystalline
order developing in the plane perpendicular to the magnetic field. 
Comparing the structure factors above and below the transition,  
the difference is most significant for the case of $n=2$. This 
tells us 
that the state
below the transition is a vortex liquid with larger correlation 
along the field direction.  
Therefore the transition we find is more like a decoupling 
transition occuring in the 
vortex liquid. 
This can also be seen in the drop in $\eta$ (Fig.~2) as the 
temperature is lowered, since $\eta$ is largest when  neighboring 
layers
are more or less independent. 

\begin{figure}
\centerline{\epsfxsize=7cm\epsfbox{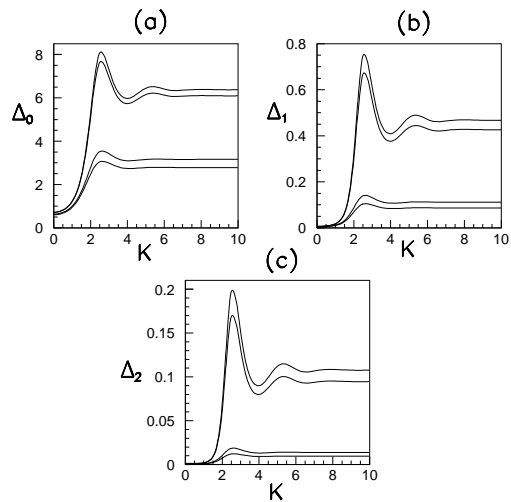}}
\vspace{10pt}
\caption{The structure factor $\Delta_n (K)$ at
$\tau_T=0.105$ and four different values of
$\alpha_T$. The dimensionless two-dimensional
wavevector $K\equiv k/\mu$. (a) n=0,(b) n=1, and
(c) n=2. For all cases, $\alpha_T$= -3.715, -3.709,
-3.685 and -3.566 reading from top to bottom.
}
\end{figure}

Figure 5 shows a collection of the first-order transition 
points in the $\alpha_T$-$\tau_T$
plane and the location of the critical end-point. 
The transition line is almost independent of
$\alpha_T$. Although we were only able to study 
the region  close to the critical end-point, 
the nearly $\alpha_T$-independent transition
line is consistent with the result obtained in Ref.~\cite{km}.
We found it very hard to extend this line to the 
small-$\tau_T$ region. Since
the size of the jump gets bigger as we move away from the critical 
end-point
to the small-$\tau_T$ region,
a solution below the transition is going to be very different 
from the one above, which 
makes the parquet equations hard to solve by iteration.
According to Eq.~(\ref{attt}) the transition line which ends 
at large $\tau_T$ corresponds to  a critical end-point
at low magnetic field. In the H-T space,
the shape of the phase boundary looks like the one 
found in Ref.~\cite{km},
which is  consistent with the experimental 
results on YBCO-type superconductors. However, we do not expect 
that the
actual position of the transition line in our work
can be directly compared with  experiment since numerical
difficulties have limited us to studying only a small number of 
layers.  The
transition temperatures ($\alpha_T\sim -3.7$) in our model are
higher than those found in Ref.~\cite{km} 
($\alpha_T\sim -7.0$), and therefore are also higher 
than the experimental values.

\begin{figure}
\centerline{\epsfxsize=7cm\epsfbox{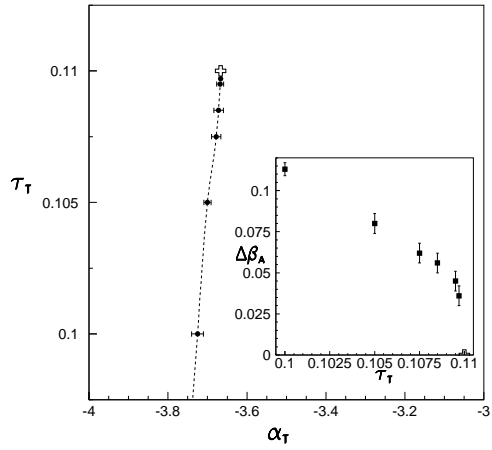}}
\vspace{10pt}
\caption{The first order transition line and the location
of the critical end-point (cross) in the $\alpha_T$-$\tau_T$
space. The inset shows how the size of the jump in $\beta_A$ at
the transition decreases as the critical end-point is approached.
}
\end{figure}

To summarize we have applied a nonperturbative analytic
method to the vortex liquid system in a layered superconductor.
The inter-layer coupling  produces
a first-order transition line which ends at
a critical end-point at low fields, whereas for a purely 
two-dimensional
system there are no transitions of any kind within the parquet 
approximation.
 The results
are consistent with the  
first-order transition being
a decoupling transition between two vortex liquid phases.
Clearly in order to extend this method to  other situations eg. 
lower 
temperatures, the effects of disorder and above all, more layers,
we have to devise a way to speed up the convergence 
of the iterative solution to the parquet equations.
One possible method might be to use a combination of solutions 
obtained in
previous steps as the next stage of the iteration \cite{ng}.

\acknowledgments

J.\ Y.\ was supported 
by the Korea Research Foundation Grant(KRF-1999-015-DI0070).


\begin{references}
\bibitem{blatter} G.\ Blatter {\it et al.}, Rev.\ Mod.\ Phys.\
{\bf 66}, 1125 (1994).
\bibitem{exp}
W.\ K.\ Kwok, J.\ Fendrich, S.\ Fleshler,
U.\ Welp, J.\ Downey, and G.\ W.\ Crabtree,
Phys.\ Rev.\ Lett.\ {\bf 72}, 1092 (1994).
\bibitem{exp1}
R.\ Liang, D.\ A.\ Bonn, and W.\ N.\ Hardy, 
Phys.\ Rev.\ Lett.\ {\bf 76}, 853 (1996).
\bibitem{exp2} A.\ Schilling, R.\ A.\ Fisher, N.\ E.\
Philips, U.\ Welp, D.\ Dasgupta, W.\ K.\ Kwok,
and G.\ W.\ Crabtree, Nature (London) {\bf 382}, 791 (1996).
\bibitem{exp3} A.\ Junod, M.\ Roulin, J.-Y.\ Genoud, B.\ Revaz,
A.\ Erb, and E.\ Walker, Physica C {\bf 275}, 245 (1997).
\bibitem{exp4} M.\ Roulin, A.\ Junod, A.\ Erb, and E.\ Walker,
Phys.\ Rev.\ Lett.\ {\bf 80}, 1722 (1998).
\bibitem{exp5} A.\ Schilling, R.\ A.\ Fisher,
N.\ E.\ Philips, U.\ Welp, W.\ K.\ Kwok, and G.\ W.\
Crabtree, Phys.\ Rev.\ Lett.\ {\bf 78}, 4833 (1997). 
\bibitem{km} A.\ K.\ Kienappel and M.\ A.\ Moore, Phys.\ Rev.\ B 
{\bf 60}, 6795 (1999).
\bibitem{period} J.\ Hu and A.\ H.\ MacDonald, Phys.\ Rev.\ B {\bf 
56},
2788 (1997); R.\ \u{S}\'{a}\u{s}ik and D.\ Stroud,
Phys.\ Rev.\ Lett.\ {\bf 75}, 2582 (1995). 
\bibitem{ym} J.\ Yeo and M.\ A.\ Moore, Phys.\ Rev.\ Lett.\ {\bf
76}, 1142 (1996); Phys.\ Rev.\ B {\bf 54}, 4218 (1996).
\bibitem{ng} K.-C Ng, J.\ Chem.\ Phys.\ {\bf 61}, 2680 (1974).
\end{references}
\end{document}